\documentclass[usegraphix,usenatbib]{mn2e}
\usepackage{natbib}
\usepackage{amsmath}
\usepackage{amssymb}
\usepackage{times}
\usepackage{graphicx}
\usepackage{subfigure}
\usepackage{txfonts}
\usepackage{aas_macros}
\usepackage{lscape}
\usepackage{color}
\usepackage{epsfig}

\begin{document}

\def\data{{\rm data}}
\def\model{{\rm model}}

\pagenumbering{arabic}

\title[Dipping Our Toes in the Water: First Models of GD-1 as a Stream]
  {Dipping Our Toes in the Water: First Models of GD-1 as a Stream}

\author[Bowden, Belokurov \& Evans]
  {A. Bowden$^1$\thanks{E-mail:adb61,vasily,nwe@ast.cam.ac.uk}, V. Belokurov$^1$, N.W. Evans$^1$ \medskip
  \\$^1$Institute of Astronomy, University of Cambridge, Madingley Road, Cambridge, CB3 0HA, UK
  \\
  }

\maketitle

\begin{abstract}
 We present a model for producing tidal streams from disrupting
 progenitors in arbitrary potentials, utilizing the idea that the
 majority of stars escape from the progenitor's two Lagrange
 points. The method involves releasing test particles at the Lagrange
 points as the satellite orbits the host and dynamically evolving them
 in the potential of both host and progenitor. The method is
 sufficiently fast to allow large-dimensional parameter exploration
 using Monte Carlo methods. We provide the first direct modelling of
 6-D stream observations -- assuming a stream rather
 than an orbit -- by applying our methods to GD-1.  This is a
 kinematically cold stream spanning $60^{\circ}$ of the sky and
 residing in the outer Galaxy $\approx 15$ kpc distant from the
 centre. We assume the stream moves in a flattened logarithmic
 potential characterised by an asymptotic circular velocity $v_0$ and
 a flattening $q$. We recover values of normalisation $v_0$ =
 $227.2^{+15.6}_{-18.2}$ kms$^{-1}$ and flattening $q$ =
 $0.91^{+0.04}_{-0.1}$, if the stream is assumed to leading, and $v_0$
 = $226.5^{+17.9}_{-17.0}$ kms$^{-1}$, $q$ = $0.90^{+0.05}_{-0.09}$,
 if it is assumed to be trailing.  This can be compared to the values
 $v_0 = 224 \pm 13$ kms$^{-1}$ and $q= 0.87^{+0.07}_{-0.04}$ obtained
 by Koposov et al (2010) using the simpler technique of orbit
 fitting. Although there are differences between stream and orbit
 fitting, we conclude that orbit fitting can provide accurate results
 given the current quality of the data, at least for this
 kinematically cold stream in this logarithmic model of the Galaxy.
\end{abstract}


\section{Introduction}

As satellite galaxies and globular clusters orbit their host galaxy,
they experience strong tidal forces. Stars are stripped from the
satellite onto orbits that are close to that of the progenitor,
forming tidal streams (e.g., Johnston et al. 1995, Fellhauer et
al. 2006). Such streams contain information about the potential in
which they form, allowing us to probe the structure of our Galaxy.

The Sagittarius (Sgr) dwarf galaxy provides the most famous example of
a tidally disrupting system. Its leading and trailing tails have now
been traced across most of the sky (e.g., Koposov et al. 2012,
Belokurov et al. 2014). The Sgr stream is unusually complex, probably
because its progenitor was a dwarf irregular comparable in mass to the
Small Magellanic Cloud (Niederste-Ostholt et al. 2010). More numerous
than such thick streams are thin and wispy tidal tails, such as the
extended trail of debris from the globular cluster Pal 5 found in
Sloan Digital Sky Survey (SDSS) commissioning data (Odenkirchen et
al. 2003).  In fact, the high quality photometric data of the SDSS
proved a gold-mine for the discovery of stellar streams. Amongst the
many discoveries were the tidal tails of NGC 5466 (Belokurov et
al. 2005, Grillmair \& Johnson 2006), the Orphan Stream (Grillmair
2006, Belokurov et al. 2007), the GD-1 Stream (Grillmair \& Dionatos
2006) and the Acheron, Cocytos, Lethe and Styx streams (Grillmair
2009).  These kinematically cold streams are derived from less massive
progenitors than the Sgr, and hence are easier to understand and
model.

An appealingly simple assumption is to model the stream as an orbit in
an underlying potential (e.g., Jin \& Lynden-Bell 2007, Willett et
al. 2010, Newberg et al. 2011, Lux et al. 2013, Deg \& Widrow
2014). Koposov, Hogg \& Rix (2010, hereafter K10) used just this
method to model the GD-1 tidal stream in a flattened logarithmic
potential. At first sight, this seems a reasonable assumption,
particularly for a narrow and long stream such as GD-1.  However,
doubts were soon raised by a number of investigators. First, Eyre
\& Binney (2011) argued that streams in realistic potentials are poorly
represented by single orbits. Then, Sanders \& Binney (2013a,b) also showed that
the misalignment between a stream and an orbit can be substantial and
cautioned against the practice of orbit fitting even for narrow
streams. They developed a formalism for stream fitting in action-angle
coordinates and tested it on mock data.

There have been a number of other recent
investigations into the problem of fitting streams to mock data
extracted from simulations. Bovy (2014) provided a framework for
computing the evolution of streams in action-angle coordinates, as
well as for deriving the probability distribution for data analysis
of the stream stars.  Bonaca et al. (2014) inserted streams into a resimulation of
the Via Lactea II simulation, created mock observations and fitted
them to constrain the host galaxy mass and to estimate the
biases. Finally, Price-Whelan et al. (2014) developed a novel method
of estimating the potential using a small number of stream stars
without explicitly modelling the stream. However, none of these
authors applied their methods to actual observational data on a stream
to demonstrate unambiguously that orbit fitting gives biased
results. We rectify that omission here, by devising our own algorithm
for fitting streams to positional and velocity data and -- after
testing and validation -- we apply it to the GD-1
stream. This stream is a particularly attractive
choice for probing the Galactic potential due to its unique 6-D
dataset.

When studying the creation and evolution of tidal streams, an obvious
tool is full N-body simulations. However, if we wish to explore
parameter space for both the potential and the progenitor's orbit,
N-body methods are too computationally intensive (e.g., Fardal et
al. 2012). Our first task is to find a rapid method of producing
streams for fitting to data. Here, we take advantage of the fact that
the majority of stars are stripped from near the satellite's Lagrange
points. Once generated, clouds of stripped stars are evolved forwards
as test particles in the underlying potential of the satellite and
host galaxy. This basic idea has been invoked in a number of recent
works on tidal tail evolution (e.g., Varghese, Ibata \& Lewis 2011,
K{\" u}pper, Lane \& Heggie 2012, Gibbons, Belokurov \& Evans
2014). We use it here to repeat the work of K10 {\it without the
  simplifying assumption that the stream delineates an orbit}. We can
therefore quantify the dangers of orbit fitting in a specific and
practical case.

The paper is arranged as follows. In Section 2, we describe the
Lagrange point stripping method used to create tidal streams.  Section
3 describes how such streams can be compared to observational data. In
Section 4, we validate the method by fitting to a stream produced
using N-body simulations. Finally, in Section 5 we apply our stream
fitting algorithm to the GD-1 stream and compare with the earlier
orbit fitting results of K10.

\begin{figure}
  \begin{center}
    \includegraphics[scale=0.23]{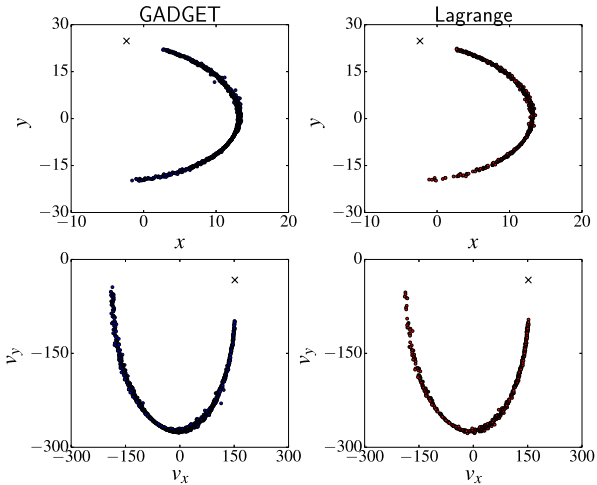}
  \end{center}
\caption{Comparison of a trailing stream produced via {\tt GADGET}
  N-body simulations to one produced with a Lagrange point stripping
  method in both position and velocity. The satellite location is
  represented by a black cross. The broad properties of the stream in
  both position and velocity space are well reproduced.}
\label{fig:streams}
\end{figure}

\section{Stream Creation by Lagrange Point Stripping}

The computational cost of N-body simulations of a tidal stream
formation lies in the calculation of the force between all the
particles within the disrupting progenitor, particularly for smaller
objects which have dynamical timescales orders of magnitudes smaller
than their orbital periods. However, we are primarily interested in
the stars which escape from the satellite and form the tails. A number
of recent studies have shown that a convenient shortcut is to follow
only the orbits of stars escaping through the two Lagrange points (see
e.g., K{\"upper et al 2012).

The method used in this paper is broadly similar to,
and was developed in conjunction with, the work of Gibbons et
al. (2014, hereafter G14), where a more detailed explanation of some
of the dynamical motivations can be found. Clouds of particles are
released from the satellite at or around the Lagrange points, and
their orbits integrated forward in the potential of both the host
galaxy and the satellite. We evaluate the Lagrange radius as
\begin{equation}
r_{\rm t} = \Biggl(\frac{GM_{\rm sat}}{\Omega^2 - \frac{d^2\Phi}{dr^2}}\Biggr)^{\frac{1}{3}},
\end{equation}
where $\Omega$ is the satellite's angular velocity and $\Phi$ is the
halo potential. For this calculation of the Lagrange points, the
satellite is approximated as a point mass. Stars are stripped from
both the inner and outer Lagrange points at Galactocentric radii
\begin{equation}
r_{\rm strip} = r_{\rm sat}-\lambda r_{\rm t} \quad{\rm or}\quad r_{\rm sat}+\lambda r_{\rm t},
\end{equation}
where $\lambda$ of order unity and $r_{\rm sat}$ is the instantaneous
position of the satellite's centre. If $\lambda = 1$ (exact Lagrange
point stripping, as used in G14), we find that a substantial number of the generated
stars are re-captured by the satellite. This is particularly
problematic just after pericentre, when the tidal radius of the
satellite is growing rapidly. As we are only interested in escaping
stars, this is computationally inefficient. For the purposes of the
model, we use a value of $\lambda$ = $1.2$, which ensures that the
overwhelming majority of the stars become part of the stream.

With the starting positions of the stripped stars in hand, they must
be assigned velocities. The base velocity is determined by the orbit
of the satellite, with the radial component matched directly to the
satellite's. For the tangential component, there are two obvious
extremes -- we can match either the satellite's velocity
(the choice made in G14) or its angular velocity.  We
find that the different choices have only a modest effect on the mean
track of the stream over a range of masses. We choose to match to the
angular velocity halfway between the satellite and the Lagrange point.
This is similar to the method described in K{\" u}pper et al.(2012),
and successfully reproduces streams from simulations.  We add a random
component to the velocities, described by a parameter $\sigma_{\rm s}$
which represents the satellite's velocity dispersion. Three components
resolved with respect to the progenitor's orbital plane $(v_{r},v_{t},
v_\perp)$ are drawn randomly from a normal distribution with zero mean
and standard deviation $\sigma_{\rm s}$.

As a satellite approaches perigalacticon, its tidal radius shrinks as
the effect of the host galaxy grows. Conversely, at apocentre the
tidal radius is at its largest. Consequently, the stripping of debris
from a disrupting satellite is not uniform in time. N-body simulations
show that whilst stars are stripped throughout the orbit, an excess is
lost near pericentre. This is one of the factors controlling the
density distribution along the stream. Others include the density
distribution of the progenitor, the epicylic motion of stars along the
stream (e.g., K{\" u}pper et al. 2012) and any possible interactions
with other structures such as dark matter subhaloes. To sidestep this
problem, we choose to fit only the mean track of the stream. This
simplifies the problem greatly, and allows us to strip stars uniformly
in time.

The orbits of the stripped stars are then integrated in the potential
of the satellite and the host galaxy, producing a model tidal
stream. Throughout this paper, we take the satellite potential to
be a Plummer model, though different functional forms can easily be
used. The algorithm used is the standard Runge-Kutta-Fehlberg method
with an adaptive step size (e.g., Press et al. 2007). Relative and
absolute error tolerances were set to $10^{-5}$, providing a
reasonable trade-off between speed and accuracy. Generally, the age
of a tidal stream is not well known, leaving us with the question of
how long the stripping process should continue. We therefore also add
a parameter $t_{\rm d}$, the disruption time, 
determining how long ago the first star was stripped.

To establish confidence in our stream generation method, Fig. ~\ref{fig:streams}
shows a comparison of streams produced using the Lagrange point
stripping model with an N-body simulation, using the freely available
tree code {\tt GADGET-2} (Springel 2005).  The model stream from
the Lagrange point stripping method is virtually indistinguishable
from the N-body stream.

In fitting, we model the data in a specific arm of the stream, and
thus we wish to avoid contamination from particles which are either in
the satellite or the wrong arm of the stream.  However, not all of the
stars stripped using this method become part of this stream. Some are
recaptured by the satellite or escape into the other arm of the
stream.  Consequently, we remove any stars which do not escape the
satellite, defined as stars which return to within a factor of $0.8$
times the tidal radius at any time during their orbit integration. In
practice, this tends to be roughly a quarter of the stars we generate.

\section{The Mechanics of Fitting}

\subsection{Preliminaries}

First, the model stream stars are converted from Cartesian coordinates
to observables, namely angular position on the sky $(\phi_1,\phi_2)$,
heliocentric distance $d$, line-of-sight velocity an $v_{\rm los}$ and
proper motion $(\mu_{\phi_1},\mu_{\phi_2})$.  For this purpose, we
assume that the Solar motion is represented by the circular orbit
given by the Galaxy potential at the Solar radius of $8.5$ kpc,
together with a correction term for the Solar peculiar motion of
$(10.0,5.25,7.17)$ kms$^{-1}$ (Dehnen \& Binney 1998). This matches
the values from K10, who used GD-1 to constrain Milky Way parameters
via an orbit fitting method.  This makes sense, as our purpose is to
make a direct comparison between the two techniques.

Second, we restrict ourselves to simulations that actually produce a
stream. The release of clouds of particles from the satellite's
Lagrange points does not guarantee a stream. Highly radial satellite
orbits, for example, produce shells and `umbrellas' (e.g., Amorisco
2014).  Similarly, individual stars with particularly unusual random
initial conditions may not become stream members. An example of this
is when stars undergo an energetic three-body
interaction with the host galaxy and progenitor satellite. Tidal
streams are formed from stars with energies and angular momenta
close to that of the progenitor satellite; certain types of
interaction lead to a dramatic change in these properties,
permitting stars to escape onto highly energetic trajectories and to
be flung to distant parts of the galaxy. We find that it is
sufficient to make a simple cut in position in order to remove such
stars; we therefore first sort the particles in the stream by their
angle along it $\phi_1$.  Having sorted the stream particles by angle,
we remove the outlying $2\%$ at either edge. This eliminates the
presence of contaminants and removes the poorly sampled edges of the
stream.

When we compare model and data, we extract a segment of the stream
which covers the angular extent of the observations. If the model does
not cover the entirety of this extent, or if there are very few stars
in this region, we can instantly reject the model. However, we do not
wish to have sharp spikes in the likelihood function where the
returned value changes instantly to zero, as this makes the parameter
space more difficult to explore. It is useful to let
the walkers know they are in the right region, even if the stream
does not quite cover the extent of the data. We therefore introduce
a weight function which allows the likelihood value to smoothly
decrease to zero. The returned log-likelihood (always negative, see
Section 3.3) is divided by a weight between zero and one (zero if
the range of the stream is not covered, one if it extends more than
$5^{\circ}$ further). The weight varies linearly within this
$5^{\circ}$ regime. We wish the model stream to cover a greater
range than our data points for two reasons. First, the extent of the
data points does not represent the extent of the observations, as
the data points come from binned photometric data (for simulated
streams, we can select a segment we know extends at least
$5^{\circ}$ further. Secondly, the stream density for the model
stream on average decreases as distance from the progenitor
increases; therefore, some of the model stream we expect to be
undetectable in observations.

\subsection{The stream track}

We chose to fit the data using the centroids of the model stream in
observable coordinates as a function of angle along the stream
$\phi_1$. This is in distinction to Sanders \& Binney (2013a,b) and
Bovy (2014), who describe methods in action-angle space of producing a
probability distribution function for model streams. We find that the
technique described here of using only the mean track of the stream is
sufficient to model data of the current quality.

The Lagrange point stripping method produces an adjustable number,
roughly between $500-1000$, of tracer particles which map the stream.  To
define the stream centroid for each of the coordinates along the
stream, we must solve the familiar problem of how best to fit a line
through a subset of points. For this purpose, we use a quadratic
spline, but questions arise such as how much (if at all) we should bin
the model stream particles and how smooth the curves we fit should be.

The eventual method was determined heuristically, after exploring a
number of possible options. We bin the model stars for two main
reasons. First, most of the parameters are observationally determined
from photometric data, which are already pixellated. Second, binning
serves to reduce the random noise or `bumpiness' in the likelihood
function. In general, a small change in the initial model parameters
should lead to a small change in the returned likelihood. A smooth
likelihood surface is easier for both gradient descent and Monte Carlo
methods to explore effectively. However, the random generation of
stars means that (even with fixed seeds) an infinitesimal parameter
change can lead to a star becoming a stream member or not. This can
cause curves fit directly to unbinned model particles to deviate
sharply and unphysically.

Of course, the trouble with binning is that information about the
curvature within each bin is necessarily lost. Some of the observed
coordinates can vary reasonably steeply across a bin in certain
circumstances. We can mitigate this by using a coordinate system
defined by the stream itself, in which this curvature is minimized. We
do this by fitting a smooth quadratic spline through the \textit{data}
for each of the observables as a function of $\phi_1$. These curves
are subtracted from the model stream's observable co-ordinates,
creating a new `stream-like' coordinate system. The model curve
fitting is then performed in these coordinates which have naturally
reduced curvature, thus minimizing any deviations caused by the curve
fitting.

The bin sizes are determined according to two criteria. We require
that there are a minimum number of 10 stars per bin. This gives us a
large enough statistical sample that we reduce the above effect, and
prevent small number statistics leading to outlying points. We also
set a minimum bin size of 0.1 radians. This avoids the oversampling of
high density regions of the stream, which can lead to unphysically
rapid oscillations in the mean track.

Once the bins have been determined, we can evaluate the mean values of
the observables, $\phi_2$, $d$, $v_{\rm los}$, $\mu_{\phi_1}$ and
$\mu_{\phi_2}$, as a function of angle along the stream. An unsmoothed
quadratic spline is then fit through the mean positions using the
inbuilt Python SciPy libraries. The spline is unsmoothed as the
binning process already provides smoothing. A quadratic spline is then
easily sufficient to describe the data's curvature.

\subsection{The likelihood value}

The next step is to devise a metric by which the simulated data can be
compared with the observations. The splines representing the model
define the stream centroid at any point along the stream. As the
curves pass through the observations, we can use the errors on the
data to provide a $\chi^2$ value for the fit. The log-likelihood
function is thus
\begin{equation}
 \ln\mathcal{L} = -\frac{\chi^2}{2} = -\sum_{i} \frac{(x_{\model, i}-x_{\data, i})^2}{2\sigma_i^2}, 
\end{equation}
where $x_i$ is each of the observables and $\sigma_i$ the associated
error.

\begin{table}
\begin{center}
 \begin{tabular}{l|l}
  Parameter & Description \\
  \hline
  $v_0,q$ & Potential parameters \\
  $R$, $z$ & Progenitor position \\
  $v_R$, $v_\phi$, $v_z$ & Progenitor velocity \\
  $M_{\rm s}$, $a_{\rm s}$, $\sigma_{\rm s}$ & Progenitor properties \\
  $t_{\rm d}$ & Disruption timescale
 \end{tabular}
\caption{Table displaying the parameters for a full stream model. We
  can fix the angular position $\theta$ without loss of generality.}
\label{tab:params}
\end{center}
\end{table}

\subsection{Parameter Exploration}

A popular method of exploring a large parameter space is using Markov
chain Monte Carlo (MCMC) methods. When given an a likelihood function,
these permit a fully probabilistic determination of the posterior
distribution. The premise is that a large number of `walkers' are
placed in some distribution within the parameter space we wish to
explore. They move around the space according to some prescription,
which in this case is a function of the locations of the other walkers
in the ensemble. The value of the likelihood function at the new
location is then compared to the original position; if the likelihood
is greater, the walker moves to the new location. If it is less, it
moves with some probability. It can be mathematically proved that
after an infinite number of steps the distribution of walkers
represents the posterior distribution. In practice, we consider the
MCMC chain to be converged once the distribution of walkers is no
longer changing significantly. As this is a probabilistic process,
appropriate priors on distributions of parameters can be included. The
code we use is a modified version of the open source {\tt EMCEE}
(Foreman-Mackey et al. 2013). The modifications allow us to store
arbitrary metadata (such as the current progenitor position for the
most likely stream candidate) from the likelihood calls.

Standard MCMC algorithms do not necessarily deal well with multi-modal
or complex likelihood contours (e.g., Feroz \& Hobson 2008).  The
primary source of this is the unavoidable discretizations in the
model. For example, we only sample a finite number of epochs, and the
model stream only contains a finite number of stars.  Another source
of bumpiness in the likelihood surface is the procedure of fitting
splines through the stars in the model stream.  Whilst we have
attempted to minimize deviations in the fitting process, we still find
that small changes in the positions of the stars can still lead to
significant changes in the spline fits. This translates to an
uncertainty in the model, causing random noise on small scales in the
returned likelihood. This in turn means small parameter variations can
lead to larger than desired changes in the likelihood value.  We find
that the {\tt EMCEE} code's built in parallel tempering sampler
helps us navigate this issue. This works via the concurrent running of
MCMC chains at higher temperatures and allowing walkers to move
between chains. These higher temperature chains have larger step sizes
and accept distant points more frequently, allowing safe navigation of
the `bumps'. This helps prevent walkers becoming stuck in local
minima.

Whilst the small-scale `bumpiness' can be dealt with, it unavoidably
leads to relatively low acceptance fractions. These are of order
$10\%$ in the lowest temperature chain. As a consequence, chains can
take a relatively large number of steps to converge.One simple test to
assess convergence is look at the mean likelihood value in the chain
on a step by step basis. If this is still changing, we can guarantee
the chain has not converged, though the converse is not true. Once the
mean likelihood value in the chain has reached a constant, we examine
convergence by assessing whether the posteriors are changing,
comparing them for two subsets of walkers in the
chain. For example, if there are $4000$ steps, and the
posteriors in the steps $3001-3500$ are the same as those within
$3501-4000$, we expect the chain to have converged. Another test of
convergence is the Gelman and Rubin (1992) diagnostic, which is a
technique based on the within-chain and between-chain variance for a
number of MCMC chains. This allows us to evaluate the potential
scale reduction factor, a statistic that is expected to be close to
unity for a converged chain. We find that for all of our sample runs
and our analysis of the GD-1 stream, this factor is within $0.5\%$
of unity ($<1.005$).

When performing our parameter exploration, we use 2048
  walkers split evenly between four temperatures. The temperature
  ladder scales exponentially, with each temperature increasing by a
  factor of $\sqrt{2}$. The {\tt EMCEE} code implements `one
  temperature swap proposal per walker per rung on the temperature
  ladder after each ensemble update'. The chains are run for a total
  of 4000 steps (over 8 million individual likelihood calls). We
  consider the first 3000 of these steps to be our burn-in period,
  leaving a final lowest temperature chain with around five hundred
  thousand steps. If there were reason to believe that the chains had
  not converged, then the burn-in period could be extended. However
  this has not proved necessary in any case thus far.

Theoretically, our complete parameter space for a full
  MCMC run consists of the potential parameters, six satellite phase
  space coordinates for the progenitor position, and four parameters
  giving the satellite mass, size, velocity dispersion and disruption
  timescale (see Table~\ref{tab:params}). We can
  however reduce this dimensionality by one by taking advantage of the
  fact that when we produce a stream, we do not only compute its final
  configuration. Instead, we can store the positions of the stars in
  the stream at a number of different snapshots. We do this at $500$
  equally spaced timesteps throughout the disruption timescale $t_{\rm
    d}$. We therefore do not require six parameters for the satellite
  phase space position and a timescale; instead, a satellite orbit can
  be defined by five varying MCMC parameters and one fixed
  co-ordinate. The likelihood function described in the previous
  section can then be evaluated at the various stored epochs. As the
  epoch varies, so evidently does the satellite position, meaning our
  `fixed' coordinate does not in fact remain the same for each
  likelihood call.

The coordinate we choose to fix is the azimuthal angle, $\theta=\theta_f$.
We choose $\theta_f$ such that the progenitor passes through
$\theta_f$ in the near future. For example, for a leading stream we
choose $\theta_f$ to be at the edge of the stream, as we know the
progenitor will imminently pass through this angle. Consequently,
when the satellite orbit is integrated backwards and the stream
produced, we can guarantee that the `correct' epoch (the satellite's
present day position) will be included.

In general, our disruption timescale consists of a
  large number of satellite angular periods, and thus the present day
  $\theta$ value of the progenitor occurs a number of times. In
  theory, we can therefore find that the stream fits the data well at
  a late snapshot (the most recent wrap of the orbit), or a much
  earlier snapshot corresponding to an earlier angular passage. This
  means that we artificially introduce a multi-modality into our MCMC
  posterior for our fixed $\theta_f$, corresponding to the remaining
  five phase space co-ordinates of the orbit. These five parameters
  are different for a fixed $\theta_f$ on subsequent wraps of the
  orbit, introducing a degeneracy despite the fact they correspond to
  the same satellite orbit and thus the same solution.  When we
  integrate backwards from these co-ordinates, we find a stream
  which fits the data regardless of whether it was on the most recent
  wrap. As multi-modal posteriors are more difficult to explore, we
  wish to prevent this happening. Consequently, we do not evaluate our
  likelihood function at every snapshot. Instead, we only extract the
  most recent `wrap', such that the range of progenitor positions
  covers no more than $2\pi$.

After the likelihood is calculated for these snapshots, we perform
a zoom-in about the best epoch, interpolating the position of each
star in the stream at 200 evenly spaced times between the epoch
prior to and immedaitely after the best epoch. The best fit of
this zoom-in is returned as the likelihood value of the fit.
Theoretically, we should in fact marginalize over all the likelihood
evaluations at the different timesteps for each evaluation in our
MCMC. However, in practice we find that the likelihood very sharply
peaked in time. Combined with the fact that the width of this peak
remains consistent for similar orbits, it can be well modelled by
a delta function, meaning taking the best value is approximately
equivalent to marginalization.

\begin{figure}
 \begin{center}
  \includegraphics[width=0.5\textwidth]{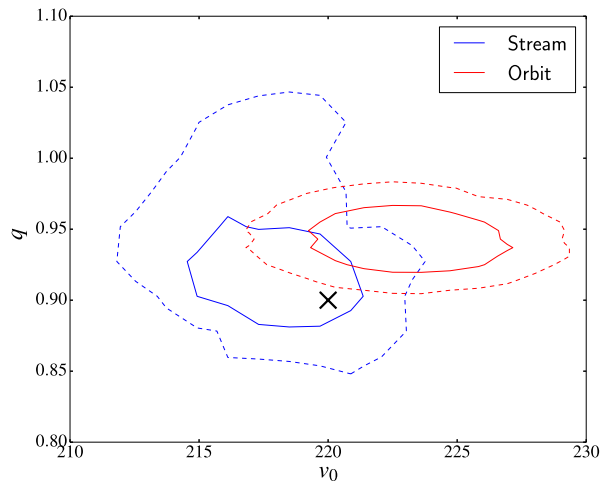}
 \end{center}
\caption{Contour plot showing the posterior probability distribution
  of potential parameters for a fit to a stream produced from an
  N-body simulation. The black cross shows the true values. The stream
  fit is shown in blue and the orbit fit in red, with the solid lines
  representing $1\sigma$ contours and the dashed lines $2\sigma$. For
  the stream fit, the recovered parameters are within $1\sigma$ of the
  correct value. The orbit fit recovers a flattening inconsistent at
  the 2$\sigma$ level with the correct value.}
\label{fig:logfit}
\end{figure}

%

\section{A Sample Fit}

In order to test the entire algorithm, we simulate a tidal stream with
{\tt GADGET} and try to recover potential parameters. For this case,
we used the correct functional form of the potential and assume
knowledge of properties (the size and mass) of the progenitor.

We produced a tidal stream via an N-body disruption of a Plummer
satellite orbiting in a fixed logarithmic potential of the form
(e.g., Binney \& Tremaine 1987)
\begin{equation}
 \Phi(R,z) = \frac{v_0^2}{2} \ln \left( R^2 + \frac{z^2}{q^2}\right),
\end{equation}
with $v_0$ = $220.0$ kms$^{-1}$ and $q$ = $0.9$. A satellite of mass
$2.5\times10^5 M_\odot$ was placed on an orbit beginning at
$(x,y,z,v_x,v_y,v_z)$ given by $(26.0,0.0,0.0,0.0,-141.8,83.1)$ in
units of kms$^{-1}$ and kpc. The positions and velocities of the
particles in the satellite were drawn from a Plummer sphere with scale
radius $a$ of 8 pc. The simulation was evolved for $10$ Gyr.

Simulation data are first converted to observational `data' for a
stream roughly 6.5 Gyrs old. We extract an angular segment of the
stream corresponding to the range we wish the fake `data' to
cover. The stars in this segment are then sorted by angle along the
stream. Their Cartesian coordinates are transformed into observables
$(\phi_1,\phi_2,d,v_{\rm los},\mu_{\phi_1},\mu_{\phi_2})$.  using the
solar position, Local Standard of Rest and solar peculiar velocity.
We then binned the stars in 10 evenly spaced bins in $\phi_1$, the
angle along the stream, and took the means of the other observables in
these bins as the data. Gaussian noise was added to each value, with
the error for each point taken as the standard deviation of the
Gaussian. These standard deviations for the coordinates
($\phi_2$/rad, $d$/kpc, $v_r$/kms$^{-1}$, $\mu_{\phi_1}$/masyr$^{-1}$, 
$\mu_{\phi_2}$/masyr$^{-1}$) were given by $(0.005,0.25,2.5,0.125,0.125)$.
With the exception of the $\phi_2$ coordinate, these values are overly
optimistic for current data.

We fit this stream using the model described in Sections 2 and 3,
extracting the potential parameters. The results of the fit are shown
in blue in Fig. ~\ref{fig:logfit}, with solid and dashed lines showing
one and two sigma contours. The recovered potential parameters are
$v_0$ = $218.0^{+2.1}_{-1.6}$ kms$^{-1}$ and $q$ =
$0.921^{+0.017}_{-0.024}$, within $1\sigma$ of the correct values as
marked by the black cross.  We can compare this to the result of orbit
fitting to the same data, shown in the red in
Fig.~\ref{fig:logfit}. The recovered potential parameters are $v_0$ =
$223.1^{+2.3}_{-2.6}$ kms$^{-1}$ and $q$ =
$0.943^{+0.012}_{-0.018}$. This plot does show the superiority of
stream fitting, as the recovered parameters are recovered more
closely.  The likelihood contours are slightly offset from the true
value due to the noise added to the fake observations of the stream.
The orbit fit is slightly worse, and the axis ratio is now no longer
consistent at the 2 $\sigma$ level with the correct value. However, it
is worth noting that the likelihood ($\chi^2$) values returned by the
two models are very similar -- the stream can be well fit by an orbit,
just not an orbit in the correct potential (which for this idealised
experiment we know).


\begin{table}
\begin{center}
 \begin{tabular}{r|r|c|l|l}
   & & Parameter & & \\
  \hline
  $5$ & $\leq$ & $R$/kpc & $\leq$ & $35$ \\
  $-30$ & $\leq$ & $z$/kpc & $\leq$ & $30$ \\
  $-100$ & $\leq$ & $v_R$/kms$^{-1}$ & $\leq$ & $200$ \\
  $-50$ & $\leq$ & $v_\phi$/kms$^{-1}$ & $\leq$ & $350$ \\
  $-200$ & $\leq$ & $v_z$/kms$^{-1}$ & $\leq$ & $0$ \\
  $130$ & $\leq$ & $v_0$/kms$^{-1}$ & $\leq$ & $290$ \\
  $0.5$ & $\leq$ & $q$ & $\leq$ & $1.5$ \\
  $4.5$ & $\leq$ & $\log(M_{\rm s}$/$M_\odot)$ & $\leq$ & $5.5$ \\
  $-3$ & $\leq$ & $\log(a_{\rm s}$/kpc) & $\leq$ & $-2$ \\
  $0.5$ & $\leq$ & $\sigma_{\rm s}$/kms$^{-1}$ & $\leq$ & $2.5$ \\
  $2$ & $\leq$ & $t_{\rm d}$/kpckm$^{-1}$s & $\leq$ & $5$ \\
 \end{tabular}
\caption{Table displaying the uniform model priors for the leading and
  trailing fits to GD-1 stream.}
\label{tab:gd1priors}
\end{center}
\end{table}

\begin{figure*}
 \begin{center}
    \includegraphics[width=0.9\textwidth]{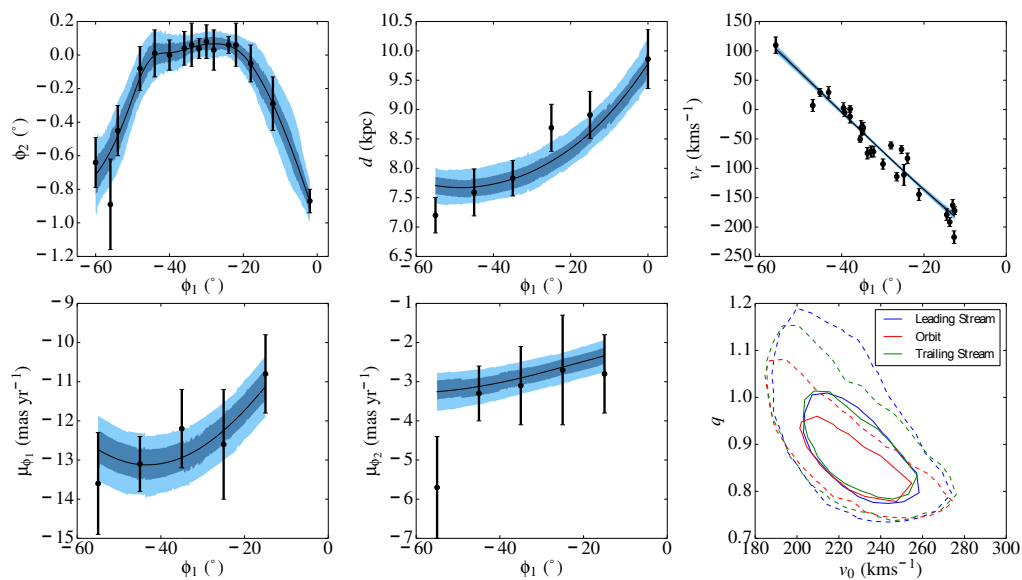}
 \end{center}
\caption{Results for fitting GD-1. The bottom right panel shows the
  recovered posterior probability distribution of potential parameters
  for the GD-1 stream. Green, blue and red lines show a trailing
  stream fit, leading stream fit and orbit fit respectively, with the
  solid lines representing $1\sigma$ contours and the dashed lines
  $2\sigma$. The remaining five panels show an
    aggregation of 10,000 randomly selected leading stream fits from
    the post burn-in MCMC chain - the trailing stream and orbit fits
  are of a very similar quality. The dark blue shaded regions cover
  $68.3\%$ of the models, and the light blue $95.5\%$. The models
  provide a good fit to the data in all observable co-ordinates.}
\label{fig:gd1fit}
\end{figure*}
\begin{figure*}
 \begin{center}
  \includegraphics[width=0.9\textwidth]{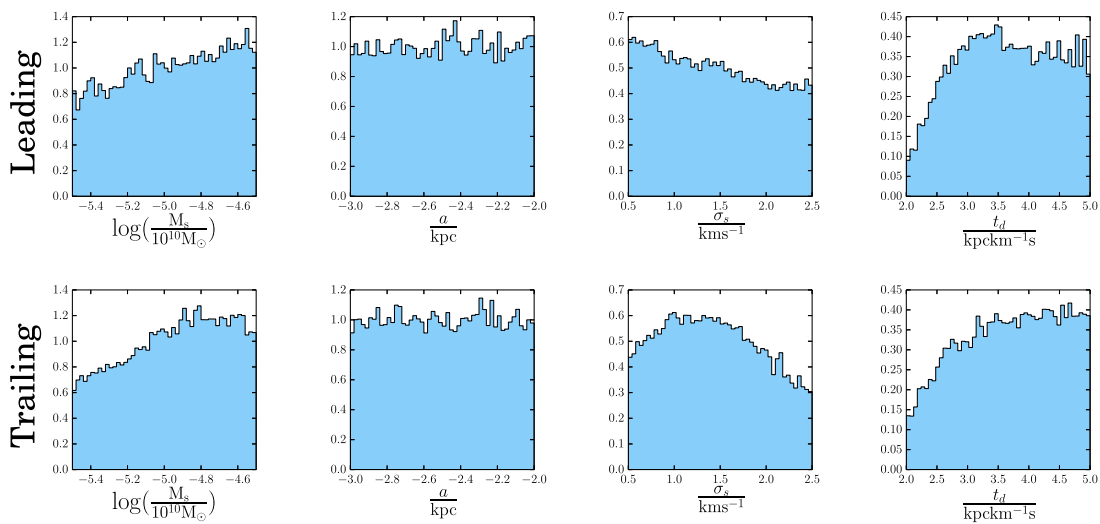}
 \end{center}
\caption{Posterior distributions on the satellite properties for
  leading (upper) and trailing (lower) stream fit to the GD-1 data. The
  only real constraining statement that can be made is that very short
  disruption times are disfavoured.}
\label{fig:satprops}
\end{figure*}

\begin{figure*}
 \begin{center}
  \includegraphics[width=0.9\textwidth]{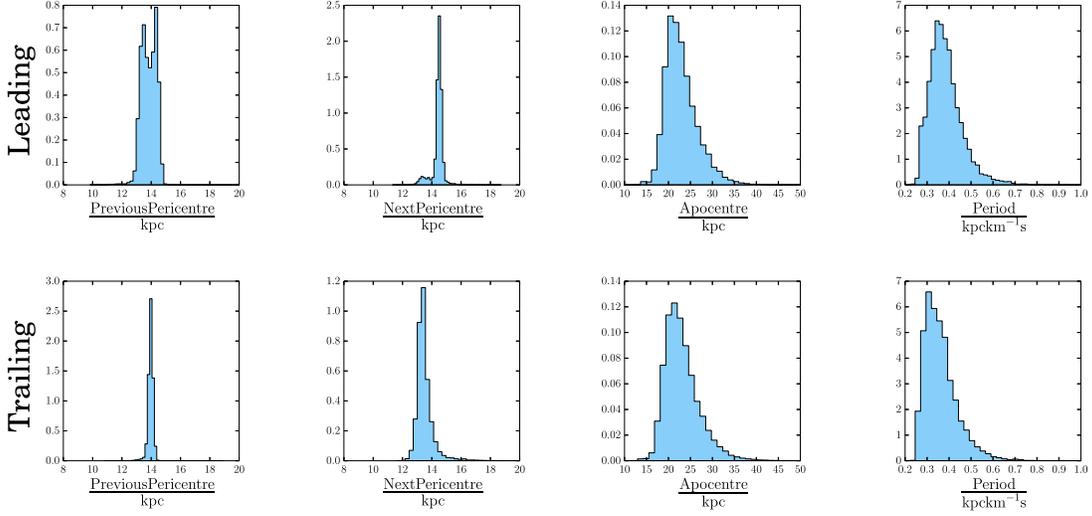}
 \end{center}
\caption{Posterior distributions on the orbital properties for the
GD-1 satellite under the assumption of a leading (upper) and trailing
(lower) stream fit. The previous and next pericentric passage distance are
displayed, along with the intervening apocentre. The period described here
is the length of time between these two pericentres; this is not a constant
value as the halo is not spherical.}
\label{fig:orbprops}
\end{figure*}

\begin{figure*}
 \begin{center}
  \includegraphics[width=0.9\textwidth]{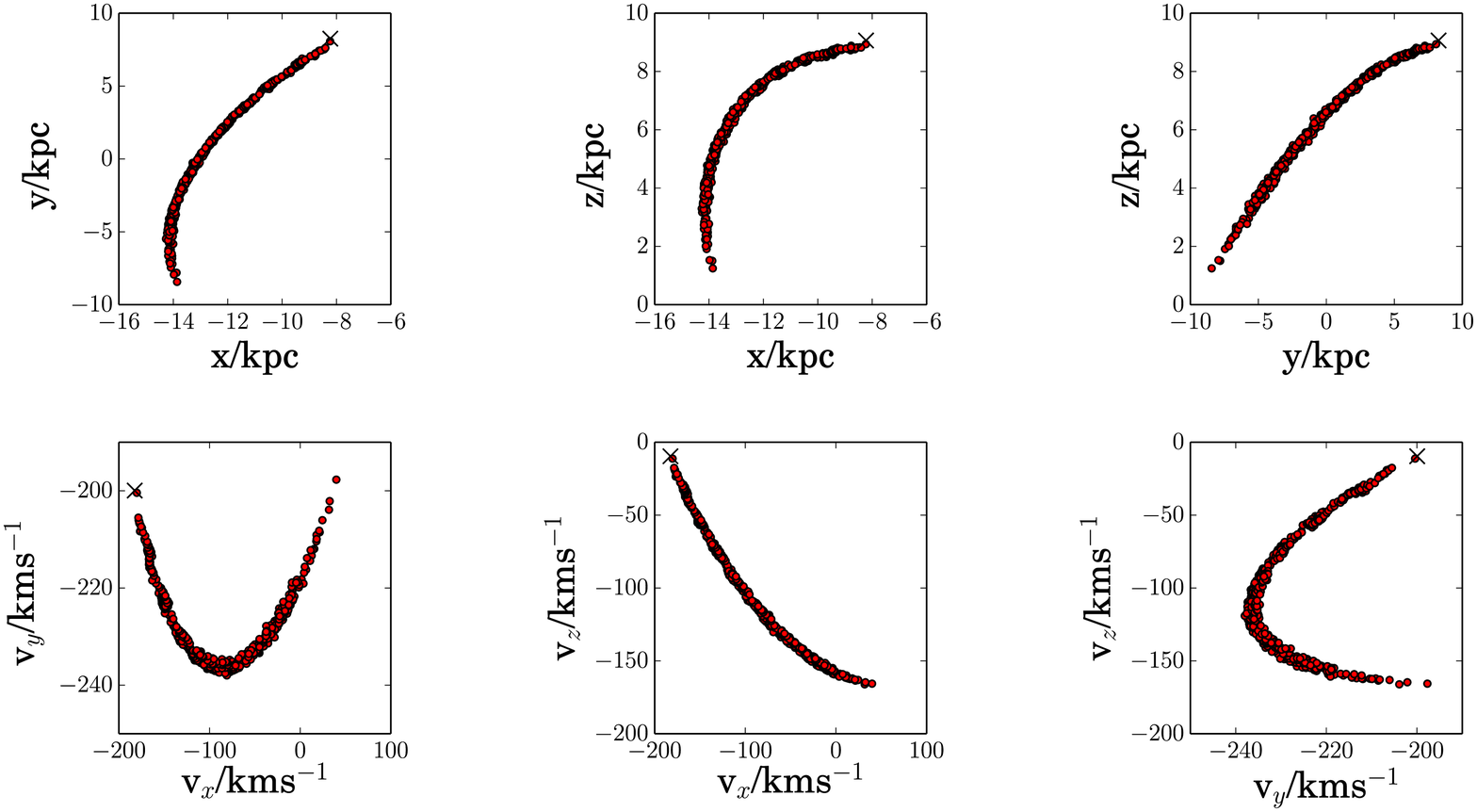}
 \end{center}
\caption{Positions and velocities for the test particles in
a model stream. The black cross shows the current position of the progenitor
in this model. The fit was for a leading stream, performed in a potential
with $v_0 = 241.6$ kms$^{-1}$, $q = 0.829$.}
\label{fig:gd1ex}
\end{figure*}

\section{The GD-1 Stream}

We now apply the algorithm described in Sections 2 and 3 to real
observations in an effort to place constraints on the potential of the
Milky Way. The stream we use is GD-1 (Grillmair \& Dionatos 2006). Its
length, spanning $60^\circ$ on the sky, as well as its relative
coldness make it an excellent target. Its biggest failing is the
limited range of distances it spans. The stream data are confined to
Galactocentric radii between 11.5 and 13.5 kpc, and to heights between
4 and 8.5 kpc above the Galactic plane.

The GD-1 data is described in full in K10, where an orbit fitting
method is used to probe the Galactic potential. It consists of 6-D
data along the stream -- positions, distances and proper motions are
determined from photometric data from SDSS. Radial velocities are from
a combination of binned SDSS data and observations of individual stars
from the Calar Alto observatory. Observations of the density
distribution along the stream are limited, so whether the stream is
leading or trailing (or both) is unknown, as is the location of the
progenitor. We will therefore perform two different fits, assuming a
leading stream and a trailing stream separately.

For the fitting, we start with wide-ranging uniform priors described
in Table~\ref{tab:gd1priors}.  The five phase space priors are
sufficiently broad to guarantee, given the properties of the stream,
that the satellite orbit is well within their range. Similarly, we can
expect the potential normalization and flattening to lie within the
bounds we have given. We cannot make the same statement for the
satellite properties, as we know little about GD-1's
progenitor. However, for a kinematically cold stream like GD-1, the
nature of the progenitor plays a relatively small role in the track of
the stream, thus we can achieve good fits with a broad range of values
for $M_{\rm s}$ and $a_{\rm s}$.  Similarly, it is difficult from the
observations to constrain the age of the stream, so we chose a
timescale for the disruption that produces a stream of roughly the
correct length.

The results of the fitting are shown in Fig.~\ref{fig:gd1fit}. The
green contours assume the data are from the trailing arm, the blue
from the leading arm, and the red contours show an orbit fit.  Solid
lines represent $1\sigma$ contours, whilst dashed lines are
$2\sigma$. The trailing stream fit recovers parameters of $v_0$ =
$226.5^{+17.9}_{-17.0}$ kms$^{-1}$, $q$ =
$0.903^{+0.049}_{-0.093}$. The leading fit recovers $v_0$ =
$227.2^{+15.6}_{-18.2}$ kms$^{-1}$, $q$ = $0.908^{+0.043}_{-0.113}$.
The orbit fit returns $v_0$ = $228.3^{+12.3}_{-12.2}$ kms$^{-1}$, $q$
= $0.856^{+0.023}_{-0.051}$.  It is interesting to note that these
values are slightly different to those recovered by the orbit fit in
K10, namely $v_0 = 224 \pm 13$ kms$^{-1}$ and $q=
0.87^{+0.07}_{-0,04}$, though they are consistent to within
$1\sigma$. This is plausibly due to the differing methodologies used
for the fitting. As visual confirmation that the fits are good, we
also show the fit to the leading stream in the space of observables in
the same figure. In Fig.~\ref{fig:gd1ex} we provide
an example of a leading stream fit with the positions and velocities
of the test particles and the present location of the satellite displayed.

Our fits do not provide strong constraints on the satellite
properties.  Posteriors for these values are shown in the panels of
Fig.~\ref{fig:satprops}.  Both fits show almost entirely flat
posteriors for the satellite mass $M_{\rm s}$ and lengthscale $a_{\rm
  s}$. There is some evidence for a slight drop off towards higher
values of the velocity dispersion $\sigma_{\rm s}$. In both cases,
the disruption time is flat towards the higher end, dropping off at
lower disruption times.

We also show our constraints on the orbital properties
  of the GD-1 progenitor in Fig.~\ref{fig:orbprops}, displaying the
  previous and next pericentre distances, the intervening apocentre
  distance, and the time between the two pericentric passages. It should be
  noted that in a flattened potential this period is not constant
  between subsequent passages. The recovered posteriors are very
  similar between leading and trailing fits, with the only discrepancy
  being the closeness of the next pericentric approach, which is
  slightly nearer the centre for a trailing stream. We observe that
  the stream is currently very near pericentre; this is as we would
  expect from its extended, thin nature. If it were near apocentre, we
  would expect more clumping of the stars. The range of apocentres and
  periods are reasonably large, spanning $~15-40$ kpc and $~250-600$
  Myr respectively. It is not clear how well constrained these
  parameters are -- the nature of the stream at pericentre is not
  necessarily affected strongly by the properties of the orbit at
  larger distances.  The recovered apocentre and period values could
  change dramatically if our assumed functional logarithmic form
  changes at larger radii.

We observe that stream fitting is consistent at the $1\sigma$ level
with orbit fitting, There are some systematic -- but slight --
offsets. Nonetheless, the agreement is impressive, especially given
the underlying simplicity of the assumption that an orbit is a
stream. Of course, this does not imply that orbit fitting is without
dangers. However, it does vindicate the work of K10
  given the quality of the current data. We have demonstrated that --
for present observations of a kinematically cold stream in the outer
parts of the Galaxy represented by an axisymmetric logarithmic
potential -- no serious problems do arise when the orbit approximation
is used.  

Sanders \& Binney (2013a) showed that a stream can only be
successfully modelled as an orbit if the misalignment angle is
small. This angle is given by
\begin{equation}
\varphi = \arccos(\hat{\boldsymbol{\Omega}}.\hat{\boldsymbol{e}}_{1}),
\end{equation}
where $\hat{\boldsymbol{\Omega}}$ is the normalised frequency vector
of the progenitor of the stellar stream and $\hat{\boldsymbol{e}}_{1}$
is the leading eigenvector of the Hessian matrix of the Hamiltonian
with respect to the actions, viz.
\begin{equation}
D_{ij}(J)=\frac{\partial^{2}H}{\partial J_{i}\partial J_{j}}.
\end{equation}
However, Hamiltonians of scale-free spherical logarithmic potentials
are linear in the actions to a very good approximation (e.g.,
Williams, Evans \& Bowden 2014) and so the misalignment angle almost
vanishes. The excellent performance of orbit fitting may therefore be
a consequence of the assumed form for the potential, and it may not
hold for more elaborate models in which the Hamiltonian has a more
complicated dependence on the actions.

The choice of a flattened logarithmic potential was primarily
motivated by the desire to perform a direct comparison with the
results of K10. Although this functional form is simple, it is a
reasonable approximation to the total Galactic potential, at least
well away from the disk.  The fitting is sensitive to the force
components along the stream. This means the constraints on the
flattening and normalization are at best valid in a narrow regime at a
distance of $\sim 15$ kpc from the Galactic Centre. Fitting GD-1 data
certainly does not constrain the global flattening, or the overall
shape or normalisation of the Galactic rotation curve,

In both leading and trailing stream cases, the recovered posteriors
are just barely consistent at the $1\sigma$ level with a spherical
potential ($q=1$). Whilst the GD-1 stream provides evidence pointing
towards a flattened potential, this is expected given its location
relative to the disk. At its closest point, the stream is $4.5$ kpc
above the Galactic plane.  Given the disk flattens the global
potential anyhow, it is evident that near-spherical halos, or even
spherical haloes, are not ruled out.  The axis ratio in the density
contours $q_\rho$ is related to the flattening in the potential by
(see eq 2.2 of Evans 1993)
\begin{equation}
q_\rho = q\sqrt{2q^2-1}
\end{equation}
The density contours are always flatter than the potential contours,
unless $q=1$. A flattening in the potential of $q=0.8$ corresponds to
an axis ration in the density of $q_\rho =0.42$, and so the $1\sigma$
contours are also consistent with a substantial range of
flattenings. However, the GD-1 stream does confirm that the halo
potential cannot be very highly flattened at these radii, which is
consistent with results from the Sagittarius stream (Evans \& Bowden
2014).

In order to approximate the effect of including a disk in our potential
model, we performed a simple test calculation. To estimate the degree
of flattening in the halo compared to the disk, we wished to compare
a flattened logarithmic halo model to a flattened log halo plus
disk. We therefore took a logarithmic halo with $v_0$ = $240.0$ kms$^{-1}$,
$q = 0.85$ and calculated its R and z derivatives across a 2-D grid in (R,z).
This grid spanned $11.5-13.5$ kpc in R and $4.5-8.5$ kpc in z,
covering the location of the GD-1 stream. We then fit these derivatives
with a disk and halo model with fixed disk parameters. Here, we took the
disk to be a razor thin exponential disk with mass $M_d$ = $6\times10^{10}
M_\odot$ and scale radius $a$ = 3 kpc. The force components for the
original potential - that which our stream model constrains - are
best fit in this case by a halo component with $v_0$ = $200.0$ kms$^{-1}$,
q = $0.866$. This suggests that such a razor thin exponential disk
has a very mild effect on the flattening of the potential at the heights
above the galactic plane occupied by GD-1.

In order to provide truly global constraints, we need to fit multiple
streams at different radii to probe more of the Galaxy.  There is
scope for the modelling and the parameter estimation that we have
developed here to play a major role.  Computational speed is not an
issue for our algorithm, with a single likelihood call generating a
stream taking of the order of a few seconds. Using modest
computational resources, this allows us to fully fit a stream on the
timescale of a week by exploring many models. For comparison, a single
N-body simulation using {\tt GADGET} carried out on a single core (as
the model is) would take of the order of a week. However, fitting
multiple streams -- possibly with other data such as the HI rotation
curve and the Oort's constants -- requires a much more flexible model
of the Galaxy's potential than has hitherto been
developed. The complicated effects of realistic halos,
  including perturbations from dark sub-halos and from nearby
  satellite galaxies, may also need to be accounted for.

\section{Conclusions}

We have carried out the first fitting of a model stream to 6-D data for
tidal stream stars in the Galaxy with an algorithm that takes into account
the fact that the stars follow a stream, and not an orbit. 
Orbit fitting has often been used before (e.g., Jin \& Lynden-Bell 2007,
Willet et al. 2010, Koposov et al. 2010, Newberg et al. 2011). In recent years,
the practice has fallen into abeyance, with a number of authors suggesting
it may give misleading or even dangerous results (e.g., Binney 2008,
Sanders \& Binney 2013a,b, Bovy 2014). However, a practical example of
the likely biases based on data on an actual stream in the Galaxy has
been missing from the literature.

Streams are produced by the stripping of stars from the Lagrange
points of disrupting satellites.  This can be routinely followed by
N-body simulations, though this is too slow for proper exploration of
parameter space. Accordingly, we use a method that quickly produces
streams via the release of stars from Lagrange points and subsequent
integration of their orbits in the potential of the Galaxy and
progenitor (c.f. Varghese et al 2011, K{\"u}pper et
al. 2012). Recently, Gibbons et al. (2014, hereafter G14) used a
very similar method to study the apocentric precession in the Sagittarius
stream and to infer the Milky Way mass. However, this paper provides the first
direct modelling of stream observations assuming they are indeed a
stream.

Our method is applied to the GD-1 stream, which was discovered by
Grillmair \& Dionatos (2006) in Sloan Digital Sky Survey data. It is a
narrow arc of $60^\circ$ on the sky at a distance of $\sim 15$ kpc
from the Galactic Centre. The observational data include positions on
the sky, heliocentric distances, radial velocities and proper motions.
The GD-1 stream has been fit before, but only under the simplifying
assumption that it follows an orbit (Koposov et al. 2010, hereafter
K10).  For this reason, we made the same assumption as KI0 as regards
the Galactic potential, taking it to be an axisymmetric logarithmic
model characterised by an overall normalisation $v_0$ and a flattening
$q$.  We can therefore directly compare the effects of stream fitting
and orbit fitting and elucidate any biases.

When GD-1 is fit as a leading stream, we recover parameters of $v_0$ =
$227.2^{+15.6}_{-18.2}$ kms$^{-1}$, $q$ =
$0.908^{+0.043}_{-0.113}$. When it is fit as a trailing stream,
recovered parameters are $v_0$ = $226.5^{+17.9}_{-17.0}$ kms$^{-1}$,
$q$ = $0.903^{+0.049}_{-0.093}$. Both these fits are consistent at the
$1\sigma$ level with the orbit fitting method of K10, who obtained
$v_0 = 224 \pm 13$ kms$^{-1}$ and $q= 0.87^{+0.07}_{-0,04}$. Although
there are differences in the posterior parameter values, they are
small. This therefore provides an explicit example
  showing that orbit fitting to streams can provide accurate results
  with current data -- at least for kinematically cold streams in the
  outer Galaxy.

There are still reasons to be cautious about orbit fitting. The
scale-free logarithmic potential used here and in K10 has a
particularly simple Hamiltonian structure (Williams, Evans \& Bowden
2014). This is relevant because the misalignment between the stream
and the orbit is controlled by the Hessian of the Hamiltonian with
respect to the actions (Sanders \& Binney 2013a). It
  is reassuring that orbit fitting works so well with the commonly
  used logarithmic potential, but it may fail in more complex
  multi-component potentials, as suggested by Sanders \& Binney
  (2013a).  The inner Galaxy is more poorly represented by scale-free
potentials, and so orbit fitting must remain highly suspect in this
regime still.

Fits to a single stream do not constrain the global Galactic matter
distribution, or the potential. What is constrained is the force
components in the small region of phase space covered by the
orbit. Fits to GD-1 do suggest that the overall potential is flattened
at distances of $\sim 15$ kpc from the Galactic Centre. The constraint
on the flattening is actually quite weak.  Spherical haloes are
consistent with the result at the 1 $\sigma$ level, especially given
the fact that some of the measured flattening in the potential is due
to the effects of the disk.  However, the $1 \sigma$ contours are also
consistent with substantial flattening in the density with axis ratios
$q_\rho \approx 0.4$.

To improve the constraints necessitates the fitting of more streams.
The fast stream production and fitting algorithm that
  we have developed, introduced here and in G14, can play an important
  role in multiple stream fitting. Before this can be made a practical
  tool to constrain the Galactic potential, two things are
  needed. First, observational data comparable to -- or better still
  superior to -- the quality and quantity available for GD-1 needs to
  be procured for many more streams. Second, the Galactic potential
  needs to be parametrised in a much more flexible way so that
  multiple constrainst from different streams can be fit
  simultaneously. We plan to return to this problem in the near
  future.

\section*{Acknowledgments}
AB thanks the Science and Technology Facilities Council (STFC) for the
award of a studentship. We thank Simon Gibbons, Sergey Koposov and
Denis Erkal for many useful comments and discussions on streams. The
referee is thanked for a careful and constructive reading of the
paper.

\bibliographystyle{mn2e}

\end{document}